\documentclass[aps, prl, twocolumn, nofootinbib, showpacs]{revtex4-1}
\usepackage{color}

\usepackage{amsmath}
\usepackage{amssymb,amsthm}
\usepackage{verbatim}
\usepackage{epsfig}
\usepackage{epstopdf}

\usepackage{amsmath, amsfonts, amssymb, amsthm, mathrsfs, amssymb, braket, bbm, bm,graphicx,times,txfonts,color,subfigure}
\usepackage{hyperref}

\newcommand{\tr}{\mathrm{Tr}}

\begin{document}

\title{Common Coherence Witnesses and Common Coherent States
}
\author{Bang-Hai Wang$^{1}$}
\email{bhwang@gdut.edu.cn}
\author{Zi-Heng Ding$^{2,3}$}
\author{Zhihao Ma$^{2,3}$}
\email{mazhihao@sjtu.edu.cn}
\author{Shao-Ming Fei$^{4,5}$}%
\email{feishm@cnu.edu.cn}
\affiliation{%
$^1$School of Computer Science and Technology, Guangdong University of Technology, Guangzhou 510006, China\\
$^2$School of Mathematical Sciences, MOE-LSC, Shanghai Jiao Tong University, Shanghai 200240, China\\
$^3$Shenzhen Institute for Quantum Science and Engineering, Southern University of Science and Technology, Shenzhen 518055, China\\
$^4$School of Mathematical Sciences, Capital Normal University, Beijing 100048, China\\
$^5$Max-Planck-Institute for Mathematics in the Sciences, 04103 Leipzig, Germany
}%

\date{\today}

\begin{abstract}
We show the properties and characterization of coherence witnesses. We show methods for constructing coherence witnesses for an arbitrary coherent state.
We investigate the problem of finding common coherence witnesses for certain class of states. We show that finitely many different witnesses $W_1, W_2, \cdots, W_n$ can detect some common coherent states if and only if $\sum_{i=1}^nt_iW_i$ is still a witnesses for any nonnegative numbers $t_i(i=1,2,\cdots,n)$. We show coherent states play the role of high-level witnesses. Thus, the common state problem is changed into the question of when different high-level witnesses (coherent states) can detect the same coherence witnesses. Moreover, we show a coherent state and its robust state have no common coherence witness and give a general way to construct optimal coherence witnesses for any comparable states.

\end{abstract}


\maketitle

\section{I. Introduction}

Originating from the fundamental superposition principle in quantum mechanics, quantum coherence \cite{Streltsov17,Hu2018} plays a crucial role in quantum
metrology \cite{Giovannetti04,Giovannetti11},  quantum algorithms \cite{Hillery 16}, nanoscale thermodynamics \cite{Aberg14,Gour15,Lostaglio15,Narasimhachar15,Francica19} and energy transportation in the biological systems \cite{Lloyd11,Huelga13,Lambert13,Romero14}. Detecting and quantifying quantum coherence, therefore, become fundamental problems in the emerging quantum areas. Numerous impressive schemes on measures of quantum coherence have been presented \cite{Baumgratz14,Yuan2015,Fang2018,Winter2016,Streltsov2015,Du2015,Qi2017,Napoli16,Bu2017,Huang2017,Jin2018,Xi2019}.

The coherence witness, inspired by entanglement witnesses, is arguably a powerful tool for coherence detection in experiments \cite{Piani16,Napoli16,Wang17,Zheng18,Ringbauer18,Nie19,Ma19} and coherence quantification in theory \cite{Wang21,Ma21}.
It directly detects any coherent states and gives rise to measures of quantum coherence without state tomography. Compared with the entanglement witness, the coherence witness has many different characteristics deserving to be investigated extensively.

Two natural questions arise that when different coherence witnesses can detect some common coherent states and when different coherent states can be
detected by some common coherence witnesses in finite-dimensional systems. Although these two similar questions related to entanglement witnesses have been well solved,
separably \cite{Wu06,Wu07,Hou10a}, the problems of common coherence witnesses and common coherent states remain unsolved.

In this paper we systematically investigate and solve the problems of common coherence witnesses and common coherent states. This paper is organized as follows. In section II, we review the concept of coherence witnesses and the methods of constructing coherence witnesses. In section III we show sufficient and necessary conditions for any given two or many coherence witnesses to be incomparable, and deal with problem of common coherence witnesses. In section IV, we characterize coherent states based on high-level witnesses and solve the problem when different coherent states can be detected by common coherence witnesses. Summary and discussions are given in section IV.


\section{II. Common coherence witnesses}

With respect to a fixed basis $\{|i\rangle\}_{i=1,2,\cdots,d}$ of the $d$-dimensional Hilbert Space $\mathcal{H}$, a state is called incoherent if it is diagonal in this basis. Denote $\mathcal{I}$ the set of incoherent states. The density operator of an arbitrary incoherent state $\delta\in \mathcal{I}$ is of the form,
\begin{equation}\label{ic}
\delta=\sum_{i=1}^d\delta_i|i\rangle\langle i|.
\end{equation}
Clearly, the set of incoherent states $\mathcal{I}$ is convex and compact. Since the set of all incoherent states is convex and compact, there must exist a hyperplane which separates a arbitrary given coherent state from the set of all incoherent states by the Hahn-Banach theorem \cite{Edwards65}. We call this hyperplane a coherence witness \cite{Piani16,Napoli16}. A coherence witness is an Hermitian operator, $W=W^\dag$, such that (i) $\text{tr}(W\delta)\geq0$ for all incoherent states $\delta\in \mathcal{I}$, and (ii) there exists a coherent state $\pi$ such that $\text{tr}(W\pi)<0$.
More precisely, an Hermitian operator $W$ on $\mathcal{H}$ is a coherence witness
if (i') its diagonal elements are all non-negative, and (ii') there is at least one
negative eigenvalue. Following the definition of incoherent states and the Hahn-Banach
theorem, we can restrict the condition (i) to $tr(W\delta)=0$ and relax (ii) to
$tr(W\pi)\neq0$ \cite{Napoli16,Ren2017,Wang21}. As coherence witnesses are hermitian quantum mechanical observables, they can be
experimentally implemented  \cite{Wang17,Zheng18,Ringbauer18,Nie19,Ma19}.


Since the density matrix of an entangled quantum state can not be diagonal, from the definition (\ref{ic}) an entangled quantum state must be a coherent state. Therefore, the entanglement witnesses are also kinds of coherence witnesses with respect to a fixed basis. We denote $\mathbf{S}$ the set of all separable states, $\mathbf{E}$ the set of all entangled states, $\mathbf{I}$ the set of all incoherent states and $\mathbf{C}$ the set of all coherent states. Fig. 1 (a) illustrates the schematic picture of the relations between entanglement and coherence. Therefore, we can construct coherence witnesses in a similar way of constructing entanglement witnesses \cite{Guhne09,Horodecki09}.

\begin{figure}[htbp]
\epsfig{file=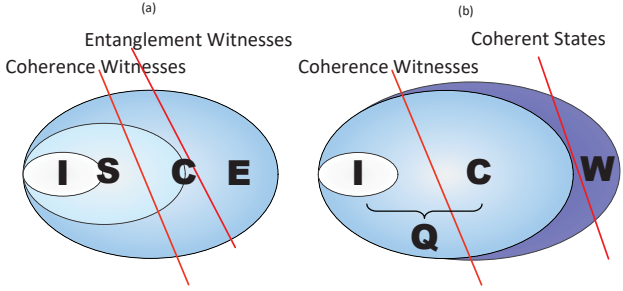,width=.95\columnwidth}
\caption{(Color online) (a) With respective to a fixed basis, all entanglement witnesses are also coherence witnesses. (b) We denote $\mathbf{Q}$ the set of all quantum states and $\mathbf{W}$ the set of all coherence witnesses. Coherent states play the role of high-level witnesses witnessing coherence witnesses.}
\label{fig1}
\end{figure}

For a given coherent state $|\psi\rangle\langle\psi|$, one has coherence witness,
\begin{equation}
W=\alpha I-|\psi\rangle\langle\psi|,
\end{equation}
where $I$ is the identity matrix and $\alpha=\max \tr(\delta|\psi\rangle\langle\psi|)$ with the maximal running over all incoherent state $\delta$.
Coherence witnesses can also be constructed from geometrical methods,
\begin{equation}
W=\frac{1}{N}(\delta-\rho+\tr(\delta(\rho-\delta))I),
\end{equation}
where $\delta$ is the closest incoherent state to $\rho$, $N=\|\rho-\delta\|$ and $\|A\|\equiv\sqrt{Tr(A\dag A)}$.
Recently, a general way of constructing a coherence witness for an arbitrary state has been provided \cite{Wang21,Ma21}:
$W_{\rho}=-\rho+\Delta(\rho)$ is an optimal coherence witness to detect the coherence of $\rho$,
where $\Delta(\rho)=\sum_{i=0}^{d-1}\langle i|\rho|i\rangle|i\rangle\langle i|$ is the dephasing operation in the reference basis $\{|i\rangle\}_{i=0}^{d-1}$.
More general constructions of coherence witnesses are also given in \cite{Wang21,Ma21}.

For a coherence witness $W$, we define $D_{W}=\{\rho\,|\,\textrm{tr}\left(\rho W\right)<0\}$, namely, the set of all coherent states `` witnessed " by $W$.
Give two coherence witnesses $W_1$ and $W_2$, we say that $W_2$ is finer than $W_1$ if $D_{W_1}\subseteq D_{W_2}$,
that is, if all the coherent states `` witnessed " by $W_1$ are also `` witnessed " by $W_2$. We call $W$ optimal if there exists no other
coherence witness which is finer than it. It is shown that a coherent witness is optimal if and only if its diagonal elements are all zero \cite{Wang21}.
For normalization we set $\|W\|_\infty=1$ as there exist traceless coherence witnesses.

Moreover, given two coherence witnesses $W_1$ and $W_2$, we say that $W_2$ and $W_1$ are incomparable if $D_{W_1}\cap D_{W_2}=\emptyset$. Two coherence witnesses $W_1$ and $W_2$ can detect some common coherent states if $D_{W_1}\cap D_{W_2}\neq\emptyset$. To proceed, we need the following lemma.

{\textbf{Lemma 1:}} If $W_{2}$ and $W_{1}$ are incomparable, i.e., $D_{W_{1}} \cap D_{W_{2}} = \emptyset $ and if $D_{W} \subset D_{W_{1}} \cup D_{W_{2}}$, then either $D_{W} \subset D_{W_{1}}$or $D_{W} \subset D_{W_{2}}$ .\par
{\it Proof.---}
On the contrary, suppose that both $D_{W_{1}} \cap D_{W}$ and $D_{W_{2}} \cap D_{W}$ are nonempty. Take $\rho_{i} \in D_{W_{i}} \cap D_{W}$ , $i = 1, 2$. Consider the segment $[\rho_{1},\rho_{2}]$ consising of
  $\rho_{t}=(1-t)\rho_{1}+t\rho_{2}$, where $0 \leq t \leq 1$.
As $D_{W}$ is convex, we obtain
\begin{equation}\label{C}
  [\rho_{1},\rho_{2}] \subset D_{W} \subset D_{W_{1}} \cup D_{W_{2}}.
\end{equation}
Thus we have
\begin{equation}\label{D}
  [\rho_{1},\rho_{2}] =( D_{W_{1}} \cap [\rho_{1},\rho_{2}]) \cup (D_{W_{2}} \cap [\rho_{1},\rho_{2}]),
\end{equation}
which means that $ [\rho_{1},\rho_{2}]$ can be divided into two convex parts.
It follows that there is $0 < t_{0} <1$ such that $\{\rho_{t}: 0 \leq t < t_{0}\} \subset D_{W_{1}}$ , $\{\rho_{t}: t_{0} < t \leq 1\} \subset D_{W_{2}}$ and either $\rho_{t_{0}} \in D_{W_{1}}$ or $\rho_{t_{0}} \in D_{W_{2}}$.\par
Assume that $\rho_{t_{0}} \in D_{W_{1}}$; then $\text{tr}(W_{1}\rho_{t_{0}} ) < 0$. Thus, for sufficiently small $\varepsilon > 0$ with $t_{0} + \varepsilon \leq 1$, we have
\begin{eqnarray*}
  0 &\leq& \text{tr}(\rho_{t_{0}+\varepsilon}W_{1}) \\
   &=& \text{tr}(\rho_{t_{0}}W_{1}) +  \varepsilon[\text{tr}(\rho_{2}W_{1}) - \text{tr}(\rho_{1}W_{1})] < 0
\end{eqnarray*}
which leads to a  contradiction. Similarly, $\rho_{t_{0}} \in D_{W_{2}}$ leads to a contradiction as well. This completes the proof.
\hfill$\blacksquare$

\textbf{ Theorem 1.}  $W_2$ and $W_1$ are incomparable (no common coherent states can be detected) if and only if there exist $a>0$ and $b>0$ such that $W_{a,b}=aW_1+bW_2$ is positive.


{\it Proof.---}
Obviously, if $W_2$ is finer than $W_1$, then $W_2$ is finer than $W_{a,b}$ and $W_{a,b}$ is finer than $W_1$ for positive $a$ and $b$.
Hence, $D_{W_1}\cap D_{W_2}\subseteq D_W=\emptyset$ since $W_{a,b}=aW_1+bW_2$ for some $a>0$ and $b>0$. Take $t=\frac{a}{b}$.\par
 By Lemma 1, we have $D_{W_{a,b}} \subset D_{W_{1}}$ or $D_{W_{a,b}} \subset D_{W_{2}}$for all $a > 0$ and $b > 0$. Then $D_{W_{a,b}}=D_{\frac{1}{b}W_{a,b}}=D_{tW_{1}+W_{2}}=D_{\frac{t}{1+t}W_{1}+\frac{1}{1+t}W_{2}}$. Hence, we obtain $D_{W_{a,b}}=D_{\lambda W_{1}+(1-\lambda)W_{2}}\doteq W_{\lambda}$ by taking $\lambda=\frac{t}{1+t}$, where $\lambda \in (0,1)$. We now can consider $W_{\lambda}$ as $W_{a,b}$. When $t$ varies from $0$ to $\infty$ continuously, then $\lambda$ varies from $0$ to $1$ continuously, which means that $D_{W_{\lambda}}$ also varies form $D_{W_{2}}$ to $D_{W_{1}}$ continuously. Take $\lambda_{0} = \text{sup}\{\lambda :  D_{W_{\lambda}}\subset D_{W_{2}} \}$. \par
We claim that if $D_{W_{\lambda_{0}}} \subset D_{W_{2}}$ then there exist $0 <\varepsilon<1 - \lambda_{0}$ such that $W_{\lambda_{0} + \varepsilon}$ is a positive operator. Otherwise, if for all $0 <\varepsilon< 1 - \lambda_{0}$, $D_{W_{\lambda_{0}+\varepsilon}} \neq \emptyset$, then we have $D_{W_{\lambda_{0}}} \subset D_{W_{2}} , D_{W_{\lambda_{0}+\varepsilon}}\subset D_{W_{1}}$, and for all $\rho \in D_{W_{\lambda_{0}}}$
\begin{equation}\label{E}
  \text{tr}(W_{\lambda_{0}}\rho) < 0, \text{tr}(W_{\lambda_{0}}\rho) + \varepsilon(\text{tr}(W_{1}\rho) - \text{tr}(W_{2}\rho)) \geq 0.
\end{equation}
Note that $\text{tr}(W_{1}\rho) \geq 0$ and $\text{tr}(W_{2}\rho) < 0$, the second part of the last inequality is positive, and $\varepsilon$ is any small positive number, so the last inequality is impossible.\par
On the other hand, if $D_{W_{\lambda_{0}}} \subset D_{W_{1}}$ then there exist $0 < \varepsilon < \lambda_{0}$ such that $D_{W_{\lambda_{0} - \varepsilon}}$ is a positive operator. Otherwise, if for all $0 < \varepsilon < \lambda_{0}$, $D_{W_{\lambda_{0} - \varepsilon}} \neq \emptyset$, then we have $D_{W_{\lambda_{0}}} \subset D_{W_{1}}$ , $D_{W_{\lambda_{0} - \varepsilon}} \subset D_{W_{2}}$, and for all $\rho \in D_{W_{\lambda_{0}}}$, we have
\begin{equation}\label{F}
  \text{tr}(W_{\lambda_{0}}\rho) < 0, \text{tr}(W_{\lambda_{0}}\rho) + \varepsilon(\text{tr}(W_{2}\rho) - \text{tr}(W_{1}\rho)) \geq 0.
\end{equation}
For the similar reason of Eq.(\ref{E}), Eq.(\ref{F}) is impossible as well.\par
To sum up the previous discussion, no matter $D_{W_{\lambda_{0}}} \subset D_{W_{1}}$ or $D_{W_{\lambda_{0}}} \subset D_{W_{2}}$, there exists $\lambda \geq0$, or equivalently $t > 0$ ($a > 0$ and $b > 0$) such that $W_{\lambda}$ ($W_{a,b}$) is a positive operator, which completes the proof of the theorem.
\hfill$\blacksquare$

\textbf{ Corollary 1.}  $W_2$ and $W_1$ are not incomparable if and only if $W_{a,b}=aW_1+bW_2$ are witnesses for all $a>0$ and $b>0$.

Theorem 1 can be generalized to the case of finitely many witnesses. We have the following result.

\textbf{ Theorem 2.}  $W_1, W_2, \cdots, W_n$ are incomparable if and only if there exist $t_i>0$ $(i=1,2,\cdots,n)$ such that $W=\sum_{i=1}^nt_iW_i$ is positive.

{\it Proof.---} (i) The ``if" part. If $W=\sum_{i=1}^nt_iW_i\ge0$ for $t_i\ge0$, then $D_W=\emptyset$.
Let $S=\{W_i|1\le i\le n\}$ and the convex hull of $cov(S)=\{\sum_{i=1}^kt_iW_i|t_i\ge0,\sum_{i=1}^Kt_i=1,W_i\in S, K\in \mathbb{N}\}$.
Without loss of generality we assume that any subsect of $S$ can detect some coherent states simultaneously.
For $n=2$, Theorem 2 holds as it reduces to the Theorem 1.
Now assume that the Theorem 2 holds for $K\le n-1$. We prove that Theorem 2 holds for $K=n$.
Indeed we only need to prove the case of $n=3$. The case of arbitrary $n$ can be proved in a similar way.

By the assumption, we have
$D_{W_1}\neq\emptyset$, $D_{W_1}\cap D_{W_2}\neq\emptyset$ and $D_{W_1}\cap D_{W_3}\neq\emptyset$.
But $D_{W_1}\cap D_{W_2}\cap D_{W_3}\neq\emptyset$, that is,
$(D_{W_1}\cap D_{W_2})\cap D_{W_1}\cap D_{W_3}\neq\emptyset$.
Let $W_{b,c}=bW_2+cW_3$, where $b>0$ and $c>0$. We have
$D_{W_1}\cap D_{W_{b,c}}\subset (D_{W_1}\cap D_{W_2})\cup(D_{W_1}\cap D_{W_3})$.
Since $(D_{W_1}\cap D_{W_2})$ and $(D_{W_1}\cap D_{W_3})$ are disjoint and $D_{W_1}\cap D_{W_{b,c}}$ is convex, $D_{W_1}\cap D_{W_{b,c}}$ varies from $(D_{W_1}\cap D_{W_3})$ to $(D_{W_1}\cap D_{W_2})$, whenever $\frac{b}{c}$ varies from 0 to $\infty$.
By the similar argument to that in the proof of Theorem 1, we conclude that there exist $\frac{b_0}{c_0}>0$ such that
$D_{W_1}\cap D_{W_{b,c}}=\emptyset$. Therefore, $W=a'W_1+b'W_{b,c}=aW_1+b'bW_2+b'cW_3\ge0$ for some $a'>0$ and $b'>0$.
By induction on $n$ we complete the proof of (i).

(ii) The ``only if" part is clear. If $D_W=\emptyset$, then there exist $W$ such that $W\ge0$ $(W\in cov(S))$ from the proof in (i).
It follows that $W$ is not a witness, which gives a contraction. \hfill$\blacksquare$

\section{III. Common coherent states}

A framework which assembles hierarchies of ``witnesses" has been proposed in \cite{Wang18}. In this framework, a coherence witness can witness coherent states, and on the other hand, a coherent state can also act as a ``high-level-witness " of coherence witnesses which witnesses coherence witnesses.
Concretely, when a coherence witness $W$ detects a coherent state $\rho$, we say that $W$ ``witnesses" the coherence of the state $\rho$.
A question naturally arises. What ``witnesses" coherence witnesses. It is known that the set of quantum states (incoherent states and coherent states) is also convex and compact.
Thus, by the Hahn-Banach theorem, there is at least one ``high-level" witness ``witnessing" a coherence witness, see Fig. 1 (b).

For a high-level witness of coherence witnesses $\Pi$, one has (i'') $\text{tr}(\Pi\varrho)\geq0$ for all quantum states $\varrho$, and (ii'') there exists at least one coherence witness $W$ such that $\text{tr}(\Pi W)<0$. Coherence witnesses ``witness'' coherent states and coherent states ``witness''coherence witnesses. Coherent states play the role of witnesses.
Since coherent states are also (high-level) witnesses, the question when different coherent states can be detected by some common coherence witnesses can be transformed into the question when different high-level witnesses (coherent states) can detect the same coherence witnesses. From the high-level-witness role played by coherent states and the Theorem 1, we have the following result.

\textbf{Theorem 3.} Two coherent states $\rho_1$ and $\rho_2$ are incomparable, i.e., $D_{\rho_1}\cap D_{\rho_2}=\emptyset$, if and only if there exists $0<t<1$ such that $\rho_t=t\rho_1+(1-t)\rho_2$ is an incoherent state.

The robust of coherence $\mathscr{C}_\mathcal{R}(\rho)$ \cite{Piani16,Napoli16} of a coherent state $\rho \in {\mathcal{D}}(\mathbb{C}^d)$ is defined as
\begin{equation}\label{ROC}
 \mathscr{C}_\mathcal{R}(\rho)= \min_{\tau \in {\mathscr{D}}(\mathbb{C}^d)} \left\{ s\geq 0\ \Big\vert\ \frac{\rho + s\ \tau}{1+s} =\delta \in {\mathcal{I}}\right\},
\end{equation}
where ${\mathscr{D}}(\mathbb{C}^d)$ stands for the convex set of density operators acting
on a $d$-dimensional Hilbert space. We have the following conclusions.

\textbf{Corollary 2:} Any coherent state $\rho$ and the state minimizing $s$ in (\ref{ROC})) $\tau$ have no common coherence witnesses.

\textbf{ Corollary 3.} Two coherent states $\rho_1$ and $\rho_2$ are not incomparable if and only if there does not exist $0<t<1$ such that $\rho_t=t\rho_1+(1-t)\rho_2$ is an incoherent state.

From the general construction of optimal coherence witnesses for an arbitrary coherent state \cite{Wang21,Ma21} and Corollary 3, there also exists a general way of constructing a common optimal coherence witness for different coherent states.

\textbf{ Corollary 4.} For two given not incomparable coherent states $\rho_1$ and $\rho_2$, the optimal coherence witness $W=aW_{\rho_1}+bW_{\rho_2}$ detects both the coherence of $\rho_1$ and $\rho_2$, where $a>0$, $b>0$ and $W_{\rho_i}=-\rho_i+\Delta(\rho_i)$ $(i=1,2)$.

It is also not difficult to generalize Theorem 3 to the case for finitely many coherent states.

\textbf{ Theorem 4.} The coherent states $\rho_1, \rho_2, \cdots, \rho_n$ are incomparable if and only if there exist $\sum_{i=1}^nt_i=1$, $t_i>0$ $(i=1,2,\cdots,n)$ such that $\rho=\sum_{i=1}^nt_i\rho_i$ is an incoherent state.


\section{IV. Summary and Discussions}

To summarize, we have investigated the properties of coherent witnesses and the methods of constructing coherence witnesses for any arbitrarily given coherent states. We have presented the conditions for different witnesses to detect the same coherent states, as well as the conditions for a set of different coherent states whose coherence can be detected by a common set of coherence witnesses. Here, we mainly considered the case of discrete quantum systems in finite-dimensional Hilbert spaces. In fact our results hold also for infinite-dimensional cases, since our main results are proved without the additional assumption $tr(W_1)=tr(W_2)$. However, the coherence in continuous variable systems (such as light modes) is significantly different from the case of the discrete systems. For instance, the set of Gaussian states must be closed and convex, but not necessarily bounded by the Hahn-Banach theorem \cite{Bruss02}). Our investigations may highlight further researches on these related problems.

\section{ACKNOWLEDGMENTS}

This work is supported by the National Natural Science Foundation of China under Grant Nos. 62072119, 61672007 and 12075159, Guangdong Basic and Applied Basic Research Foundation under Grant No. 2020A1515011180, Shenzhen Institute for Quantum Science and Engineering, Southern University of Science and
Technology (Grant Nos. SIQSE202005, SIQSE202001), Natural Science Foundation of Shanghai (Grant No. 20ZR1426400), Beijing Natural Science Foundation (Z190005), the Academician Innovation Platform of Hainan Province, and Academy for Multidisciplinary Studies, Capital Normal University.





\end{document}